# Lattice Dynamics and High Pressure Phase Stability of Zircon Structured Natural Silicates

Preyoshi P. Bose[1], R. Mittal[1,2] and S. L. Chaplot[1]

[1]Solid State Physics Division, Bhabha Atomic Research Centre, Trombay, Mumbai, 400085, India

[2]Juelich Centre for Neutron Science, IFF, Forschungszentrum Juelich, Outstation at FRM II, Lichtenbergstr. 1, D-85747 Garching, Germany

## Abstract

We report a lattice dynamics study of relative stability of various phases of natural silicates $MSiO_4$ (M=Zr, Hf, Th, U) as a function of pressure (P) and temperature (T), which is important in the context of their use in nuclear waste storage media. Extending our previous work on $ZrSiO_4$, the Gibbs free energy has been calculated using a transferable interatomic potential in various phases over a range of P and T. Due to an interesting interplay between the vibrational entropy and atomic packing, the zircon (body centered tetragonal, $I4_1/amd$), scheelite (body centered tetragonal, $I4_1/a$) and huttonite (monoclinic, $P2_1/n$) phases occur at different P and T. It is shown that for $ThSiO_4$ at high P, the huttonite and scheelite phases are favored at high and low T respectively. However, for both $USiO_4$ and $HfSiO_4$ the huttonite phase is dynamically unstable and the scheelite phase is stable as the high pressure phase. In fact, the calculations reveal that the stability of the huttonite phase is determined by the ionic size of the M-atom; this phase is unstable for the silicate with the smaller Hf and U ions and stable with the larger Th ion. The calculated phase diagrams are in fair agreement with the reported experimental observations. The calculated structures, phonon spectra, and various thermodynamic properties also fairly well reproduce the available experimental data. The low-energy phonons in the huttonite phase that contribute to its large vibrational entropy are found to involve librational motion of the silicate tetrahedral units.

**PACS numbers**: 63.20.-e, 65.40.–b

**Keywords**: Phonons, Thermal properties of crystalline solids

## I. Introduction

ZrSiO$_4$, HfSiO$_4$, ThSiO$_4$ and USiO$_4$ form the orthosilicates group of isomorphic crystals. These crystals have the zircon structure (Fig. 1) with the space group *I4$_1$/amd* (D$_{4h}^{19}$) and four formula units in the tetragonal unit cell. The structure is common to a variety of optical materials, including rare earth orthophosphates (RPO$_4$, R=Rare earth atom), vanadates (RVO$_4$) and arsenates (RAsO$_4$). High melting temperature, chemical stability and long term corrosion resistance has prompted the use of these compounds in nuclear waste storage media.[1]. These compounds in general have good optical quality, high hardness and large refractive index. In addition to this, hafnon is a candidate for replacing SiO$_2$ as a gate in the CMOS devices. Zirconium, hafnium, thorium and uranium are localized in the earth's crust during the later stages of magmatic activity and crystallize primarily as orthosilicates or oxides. Thorite exists in the monoclinic form in nature as found by Hutton, from the sands of Gillespie's beach and named as huttonite. Coffinite is found in nature with some (OH) substituting the (SiO$_4$) group. Coffinite (USiO$_4$) which is isostructural to zircon, is one of the mineral phases determining uranium solubility in accidental corrosion of nuclear fuel by geological ground water.

The study of orthosilicates, zircon (ZrSiO$_4$), hafnon (HfSiO$_4$) and thorite (ThSiO$_4$) are of particularly importance, since these compounds are effective radiation resistant materials suitable for fission reactor applications and for storage of nuclear waste.[2] The waste has to be stored under a certain temperature and pressure so as to avoid decomposition of compound. At higher temperatures these silicates decompose[3,4] into their constituent oxides and radioactive waste may distribute itself among the component oxides. In order to study the behaviour under the natural condition of temperature and pressure we have undertaken a theoretical study based on a potential model developed for zircon validated using our extensive measurements of the phonon dispersion relation and density of states.[5-8] The model is further extended to study the thermodynamic properties of the remaining orthosilicates of the type MSiO$_4$ (M=Hf, Th, U). The thermodynamic properties of the above mentioned orthosilicates compounds are not very well studied yet. Therefore, the study of macroscopic thermodynamic properties through the study of microscopic phonon behaviour in the bulk of the compound will help in understanding the behavior of these compounds under natural radiation and temperature pressure conditions prevalent under the earth's crust. In the present study since Zr and Hf is expected to have similar properties due to chemical homology, their corresponding silicates form a group. Similarly, Th and U silicates are put into another group considering the chemical homology between them. The task for prediction of high pressure phase for USiO$_4$ is simplified and achieved due to the above consideration.

Light scattering studies have been reported to measure zone centre phonon modes in zircon phase of ZrSiO$_4$, HfSiO$_4$ and ThSiO$_4$.[9-14] Density functional calculations have been carried out[15] to investigate the structural, vibrational phonon modes and dielectric properties of zirconium and hafnium silicates in the zircon phase at zero pressure. ZrSiO$_4$ and HfSiO$_4$ are known to transform[16,17] to the scheelite phase (body centered tetragonal, *I4$_1$/a*) (Fig. 1) at high pressure and temperature. Scheelite phase of ZrSiO$_4$ and HfSiO$_4$ is known as one of the most incompressible compounds containing SiO$_4$ tetrahedra. At high pressure and temperature[18] zircon phase of ThSiO$_4$ transforms into huttonite phase (monoclinic, *P2$_1$/n*) (Fig. 1). Zircon to huttonite transition is unusual since a less dense phase usually occurs at high temperature. To our knowledge there are no high pressure and temperature studies reported for USiO$_4$. For the sake of completion of the set of orthosilicates, we report the calculation based on our model for uranium silicate and also predict its high temperature and pressure phase.

Earlier we studied ZrSiO$_4$, both experimentally as well as theoretically.[5-8] The thermodynamic properties of the rest three compounds are not very well known. The interatomic potential model earlier developed[7] for ZrSiO$_4$ is now extended to HfSiO$_4$, ThSiO$_4$ and USiO$_4$. We have calculated high pressure and temperature thermodynamic properties as well as high pressure phase transformations of these compounds. The paper is outlined as follows: lattice dynamical calculations section II followed by result and discussion and conclusion in sections III and IV respectively.

**II. Lattice dynamical calculations**

The present lattice dynamics calculations involve semi-empirical interatomic potentials of the following form[7] consisting of Coulombic and short-ranged terms;

$$V(r) = \frac{e^2}{4\pi\varepsilon_0} \frac{Z(k)Z(k')}{r} + a \exp(\frac{-br}{R(k)+R(k')}) - D \exp[\frac{-n(r-r_0)^2}{2r}] \qquad (1)$$

where, "r" is the separation between the atoms of type k and k'. R(k) and Z(k) refer to radius and charge parameters of the atom of type k respectively. a=1822 eV and b=12.364. This choice was successfully used earlier to study the lattice dynamics and thermodynamic properties of several complex solids.[19-21] This procedure is found to be useful to limit the total number of variable parameters. The bond stretching potential, given by the third term is included between the Si-O bonds. D and n are the empirical parameters[7] of covalent potential and $r_o$ = 1.627 Å is Si-O bond length. V(r) in eq. (1) represents only one pair of atoms. The total crystal potential includes a sum over all pairs of atoms. The polarizability of the oxygen atoms has been included in the framework of shell model.[22]

The parameters of the empirical potential in Eq. (1) were determined such that the zircon crystal structure obtained from the minimization of free energy at T = 0 is close to that determined using diffraction experiments. The potential also satisfies the dynamical equilibrium conditions of the zircon crystal, that is, the calculated phonon frequencies have real values for all the wave vectors in the Brillouin zone. The parameters of potentials also fitted to reproduce various other available experimental data, namely, elastic constants, optical phonon frequencies or the range of phonon spectrum, etc. The crystal structures at high pressures are calculated by minimization of the free energy at T = 0 with respect to the lattice parameters and the atomic positions. The vibrational contribution was not included to derive the structure as a function of pressure. We expect a small contribution from the quantum mechanical zero-point vibrations that we have ignored. The equilibrium structures thus obtained are used in lattice-dynamics calculations. The potential reproduces the experimental data[25-27] of lattice constants and fractional atomic coordinates (Table I) of $MSiO_4$ quite satisfactorily. The good agreement between the calculated and experimental structures as well as other dynamical properties (as discussed later) indicates that our interatomic potential model for $MSiO_4$ is quite good.

We used the potential parameters[7] for $ZrSiO_4$ as the starting point for calculations of $MSiO_4$ and changed only the radius parameters associated with the M(=Hf, Th, U) atoms. The radius parameter in eq. (1) is related to the ionic radius of atom. The radius parameter for Hf-atom is obtained by scaling the radius parameter of Zr atom as determined for $ZrSiO_4$ potential[7] in the ratio of ionic radii of Hf and Zr atoms in the octahedral co-ordiation. It turns out from our calculations that the nature of phase diagram varies systematically with the radius parameter of the M-atom. The radius parameter of the Th-atom was fine tuned to reproduce the zircon-huttonite phase boundary as known from experiments. The value for the U-atom was then scaled with its ionic radius. The radii parameters used in our calculations are R(Hf)= 1.91 Å, R(Th)= 2.22 Å and R(U)= 2.11 Å. The code[23] "DISPR" developed at Trombay is used for the calculation of phonon dispersion relation, the polarization vector of the phonons, the frequency distribution of phonons, equation of state, specific heat, etc. The code DISPR uses the lattice dynamics methods described in Ref. [22] for ionic solids. The same code was previously used for similar calculations of several complex solids.[19-21]

The phase diagram of a compound can be calculated by comparing the Gibbs free energies in various phases. In quasiharmonic approximation, Gibbs free energy of $n^{th}$ phase is given by

$$G = \Phi_n + PV_n - TS_n \qquad (2)$$

Where, $\Phi_n$, $V_n$ and $S_n$ refers to the internal energy, lattice volume and the vibrational entropy of the $n^{th}$ phase. The vibrational contribution is included by calculating the phonon density of states in all the phases of $MSiO_4$ to derive the free energy as a function of temperature at each pressure. Then the Gibbs free energy has been calculated as a function of pressure and temperature. The calculation for $HfSiO_4$ and $USiO_4$ are carried out in interval of 2 GPa while for $ThSiO_4$ the step size for calculation was 1 GPa.

### III. Results and discussion

### A. Raman and infrared modes, phonon dispersion relation and phonon density of states

The calculated phonon frequencies at the zone centre for all the compounds in the zircon phase are compared in Fig. 2. The calculations are compared with the experimental Raman data and the ab-initio calculations. The average deviation between the calculated and experimental frequencies is within 4-5%. It is interesting to see variations in frequencies of some of the modes in $MSiO_4$. The changes might be due to variation of mass of the M (Zr, Hf, U, Th) ion, volume changes are due to difference in the interatomic force constants. The volume of $ZrSiO_4$ or $HfSiO_4$ is nearly same, so the effect due to volume change would be small. The effect of the mass ratio of M ion (Hf/Zr = 1.96) is clear for the $B_{1g}(1)$ mode in which the M(Hf, Zr) atoms move significantly more than O atoms. The frequency of this mode decreases by about 40% in HfSiO$_4$ as expected from the change in mass of Hf atom. The frequencies of modes should not vary much from HfSiO$_4$ to ZrSiO$_4$ in which the M(Hf, Zr) atoms are not involved, as well as for those in which the O atoms contribute significantly more than the *M*(Hf, Zr) atoms. In most of the cases this is observed. The frequencies of the lowest A$_{2u}$ and E$_u$ modes seem to be effected by the changes in mass and as well as due to difference in the interatomic force constants.

Further we have calculated phonon dispersion relation for MSiO$_4$. In order to compare the phonon dispersion in various phases of MSiO$_4$ we have chosen a common high symmetry direction $\Lambda$. The common $\Lambda$ direction is labeled as the c-axis for zircon (*I4$_1$/amd*) and scheelite (*I4$_1$/a*) phase, while it is the b-axis for the huttonite phase (*P2$_1$/n*). The group theoretical decomposition of phonon branches along the $\Lambda$ direction in the ambient as well as high pressure phases is as follows:

Zircon phase: $6\Lambda_1 + 2\Lambda_2 + 6\Lambda_3 + 2\Lambda_4 + 10\Lambda_5$ ($\Lambda_5$ being doubly degenerate)

Scheelite phase: $8\Lambda_1 + 8\Lambda_2 + 10\Lambda_3$ ($\Lambda_3$ being doubly degenerate)

Huttonite phase: $36\Lambda_1 + 36\Lambda_2$

The calculated phonon dispersion relation for HfSiO$_4$ and ThSiO$_4$ in their ambient pressure and high pressure phases are shown in Fig. 3. The c-axis in the scheelite phase is about double in comparison of zircon phase. The Brillouin zone in the scheelite phase is half and there is a folding back of the dispersion branches from zone boundary (zircon phase) to zone centre (scheelite phase). The comparison of phonon dispersion relation in ThSiO$_4$ shows that there are several low energy optic phonon modes (Fig. 3) in the Huttonite phase of ThSiO$_4$. These low energy modes are responsible for unusual zircon to Huttonite transition in ThSiO$_4$. The numbers of modes double in the Huttonite phase due to the doubling of the primitive cell size.

We have calculated one phonon density of states and partial density of states (Fig.4) for all the silicates in the ambient pressure as well as high pressure phases. Our calculations show that in the zircon phase M atoms contribute only in the low energy range up to 40 meV. The vibrations of oxygen and silicon atoms span the entire 0-135 meV range. Above 105 meV the contributions are mainly due to Si-O stretching modes. The phonon spectra in the ambient pressure i.e., the zircon phase extends upto 135 meV, while in high pressure phases the spectra softens to lower energies upto 130 meV. The softening in the spectra is due to decrease in the contributions from the Si and O. The phonon density of states in the huttonite phase of $ThSiO_4$ and $USiO_4$ has a low energy peak at about 10 meV, while there is no such peak in the scheelite phase of $HfSiO_4$. The low energy peak is mainly due to contributions from the M atoms. The partial density of states has been used for the calculations of neutron weighted phonon density (Fig. 5) of states via the relation

$$g^n(E) = B \sum_k \{\frac{4\pi b_k^2}{m_k}\} g_k(E) \qquad (3)$$

Where B is a normalization constant, and $b_k$, $m_k$, and $g_k(E)$ are, respectively, the neutron scattering length, mass, and partial density of states of the $k^{th}$ atom in the unit cell. Typical weighting factors $\frac{4\pi b_k^2}{m_k}$ for the various atoms in the units of barns/amu are: Hf: 0.057; Th: 0.058; U: 0.037; Si: 0.077 and O: 0.265 barn/amu. The values of neutron scattering lengths for various atoms can be found from Ref. [24]. Fig. 5 shows the comparison of the calculated neutron-cross-section-weighted phonon density of states in $MSiO_4$ (M=Hf, Th, U) in various phases at P=0. At present phonon density of states is not measured for these compounds. The calculations (Fig. 5) would be useful in future for comparison of our calculated phonon spectra with the experimental data.

**B. Thermodynamic properties:**

The density of states (Fig.5) is used for the calculations of specific heat (Fig. 6) in the zircon as well as high pressure phases of $ZrSiO_4$, $ThSiO_4$ and $USiO_4$. The huttonite phase of $ThSiO_4$ has higher specific heat at low temperatures in comparison of scheelite phase of $HfSiO_4$ and $USiO_4$. This is due to presence of low energy optic phonons (Figs. 3 and 4) in the huttonite phase of $ThSiO_4$ in comparison of the scheelite $HfSiO_4$ and $USiO_4$. Due to larger volume thermal expansion in the high pressure phases in comparison of the zircon phase (described below), the $C_P - C_V = \alpha_V^2 BVT$ corrections arising from anharmonicity of phonons are larger for the high pressure phases (Fig. 7).

Thermal expansion is related to the anharmonicity of lattice vibrations. In the quasiharmonic approximation, each of the phonon modes contributes to the volume thermal expansion[22(b)] equal to $\alpha_V = \frac{1}{BV}\sum_i \Gamma_i C_{Vi}(T)$. Here, $\Gamma_i$ (=-$\partial lnE_i/\partial lnV$) and $C_{Vi}$ are the mode Grüneisen parameters and the specific heat contributions respectively of the phonons in the i$^{th}$ state. Grüneisen parameters can be calculated from the volume dependence of phonon energies. The procedure for calculation of thermal expansion is valid only when the effect of explicit anharmonicity is not very significant. Due to very large Debye temperatures (~ 975 K at 1000 K) the procedure seems to be suitable up to fairly high temperatures. We have used energy dependence of Grüneisen parameter (Fig. 7) in the calculations of thermal expansion. Low energy phonon modes have large Grüneisen parameter in comparison of the high energy modes in the high pressure phases of the silicates. The calculated partial density of states shows (Fig. 4) that at low energies contributions are mainly from the M(Hf, Th, U) atoms. The calculated thermal expansion behavior is shown in Fig. 8. The large Grüneisen parameter in high pressure phase in comparison of the zircon phase results in larger thermal expansion in high pressure phase.

The crystal structures at high pressures is obtained by minimizing the Gibbs free energy with respect to the structure variables (lattice parameters and atomic positions), while keeping the space group unchanged. The calculated equation of state in the zircon as well as high pressure phases is shown in Fig. 9. For HfSiO$_4$, in the zircon phase the compressibility along a-axis is higher than that along the c-axis, while in the scheelite phase compressibility along a-axis is smaller than that along the c-axis. In the zircon phase of MSiO$_4$ compounds, the structure unit can be considered as a chain of alternating edge-sharing SiO$_4$ tetrahedra and MO$_8$ dodecahedra extending parallel to the *c*-axis, with the chain joined along the *a*-axis by edge-sharing MO$_8$ dodecahedra. The scheelite phase consists of SiO$_4$ tetrahedra aligned along the a-axis, whereas along c-axis MO$_8$ dodecahedra intersperse between the SiO$_4$ tetrahedra. At high pressure, because of the covalent nature, the Si-O bonds remain undistorted while the volume of MO$_8$ dodecahedra is reduced. This results in a smaller compressibility along a-axis in the scheelite phase in comparison of the zircon phase. The scheelite phase is less compressible in comparison of the zircon phase. The calculated equation of state for various phases (Fig. 9) of ThSiO$_4$ and USiO$_4$ show that these compounds are more compressible in comparison of HfSiO$_4$.

The computed elastic constants for the MSiO$_4$ are given in Table II. The calculated bulk modulus value of ZrSiO$_4$ is 22% higher than the experimental[30] value. However, the calculated acoustic phonon branches for ZrSiO$_4$ are found to be in good agreement with the calculations in our previous paper.[7] Therefore the bulk modulus should also be well reproduced. Perhaps the measurement of the bulk modulus of ZrSiO$_4$ from natural single crystals may have been influenced[28] by the presence of known radiation damage due to radioactive impurities. This may be one of the reasons for difference between the experimental and calculated value of bulk modulus. The calculated bulk modulus values of the zircon and scheelite phases of HfSiO$_4$ are 260 and 314 GPa respectively. These values are about 3.5 % higher in comparison of the ZrSiO$_4$. The calculated bulk moduli for ThSiO$_4$ and USiO$_4$ in their zircon phase are nearly same. These values are about 80 % of the bulk modulus values of HfSiO$_4$.

**C. Gibbs free energies and phase stability:**

The zircon structure compounds are known to transform to the scheelite phase ($I4_1/a$) at about 20 GPa. However, thorium silicate, ThSiO$_4$ has a zircon structure ($I4_1/amd$) at low temperature[25], whereas the high temperature form of ThSiO$_4$ has huttonite structure ($P2_1/n$). Zircon to huttonite transition is unusual[4,16] since a less dense phase usually occurs at high temperature. High pressure studies have not been reported for USiO$_4$. We note that for ThSiO$_4$, there is a greater density of low frequency modes (Figs. 3 and 4) in the huttonite phase in comparison of the zircon phase. This result in larger vibrational entropy in the huttonite phase, which favors this phase at high temperature.[19(b)] Figure 10 shows typical plots of the differences in the free energies of competing phases as a function of pressure or temperature.

It is found that the zircon phase transforms (Fig. 11) to the scheelite and huttonite phases at high pressure for HfSiO$_4$ and ThSiO$_4$ respectively, which is in good agreement with the experimental observations.[16,17] It is likely that the phase transition pressure of ThSiO$_4$ in experiments is overestimated as these were performed with only increasing pressure and some hysteresis is expected.[16] Our calculated transition pressure agrees with that estimated from an analysis of the measured enthalpies.[16] For ThSiO$_4$, at further high pressure, the scheelite phase is found (Fig. 11) to be stable. Experimentally, however, transformation to an amorphous phase is found (coexisting with the huttonite phase) instead of the scheelite phase, which might be due to kinetic hindrance. The free energy calculations in the zircon, scheelite and huttonite phases of $USiO_4$ suggest that scheelite is the stable phase of $USiO_4$ at high pressures. The free energy changes due to volume are important in zircon to scheelite phase transition in $HfSiO_4$ and $USiO_4$, while vibrational energy and entropy play an important role in zircon to huttonite phase transition in $ThSiO_4$. It is very important and satisfying to note that the free energy calculation with the present model is able to distinguish between the phases and reproduce their relative stability over a range of pressure and temperature. This is probably the most stringent test of the interatomic potentials.

As noted above, the greater density of low-energy modes in the huttonite phase is the key to the zircon to huttonite phase transition at high temperature. In order to understand the nature of low energy phonon modes in various phases of $ThSiO_4$ we have calculated mean squared amplitude ($<u^2>$) of various atoms at T=300 K (Fig. 12) arising from phonons of energy E integrated over the Brillouin zone. The modes up to 4 meV involve equal amplitudes of various atoms, and so are largely acoustic in nature. For 5 to 20 meV in the zircon and scheelite phases, the calculated $<u^2>$ values of Si and O atoms are nearly the same which indicates that these modes involve translation of $SiO_4$ tetrahedra as a whole. On the other hand, in the huttonite phase, significantly larger amplitude of the O atoms in comparison of the Si atoms indicates the libration of $SiO_4$ tetrahedral units in addition to the translational motion. Further, the various oxygen atoms constituting the $SiO_4$ tetrahedra in huttonite phase have significantly different values of their vibrational amplitudes, which indicate distortions of the $SiO_4$ tetrahedra. In summary, it appears that the huttonite phase has a greater density of the librational modes of the silicate tetrahedra at low energies, and that seems to be the key to the zircon to huttonite phase transition. In the zircon and the scheelite phases, the librational modes occur at much higher energies around 30 meV.

**IV. Conclusions**

We have developed a lattice dynamical shell model for various zircon structured compounds $MSiO_4$ and validated it by previous extensive measurements of the phonon density of states and phonon dispersion relation, and other data available in the literature. Further, we employed this model for calculation of various high pressure and temperature thermodynamic properties of $MSiO_4$ in their zircon, scheelite and huttonite phases. The lattice dynamical models are further used to calculate the free energies as a function of pressure and temperature in the zircon as well as the high-pressure scheelite and huttonite phases. The calculated free energies reproduce the relative stability of the phases across their observed phase transition pressure and temperature.

**References**

[1] L. A. Boatner, G. W. Beall, M. M.Abraham, C. B. Finch, P. G. Huray and M. Rappaz, in *The Scientific Basis for Nuclear Waste Management*, Vol. **II**, ed. by C. J.Northrup, p. 289. Plenum Press, New York (1980); T. Hayhurst, G. Shalimoff, N. Edelstin, L. A. Boatner and M. M. Abraham, J. Chem. Phys. **74**, 5449 (1981).

[2] R. C. Ewing et al., J. Mater. Res. **10**, 243 (1995).

[3] L. G. Liu, Earth Planet. Sc. Lett. **44**, 390 (1979).

[4] F. A. Mumpton and R. Roy, Geochimica et Cosmochimica Acta **21,** 217 (1961.

[5] R. Mittal, S. L. Chaplot, R. Parthasarathy, M. J. Bull and M. J. Harris, Phys. Rev. B **62,** 12089 (2000).

[6] R. Mittal, S. L. Chaplot, Mala N. Rao, N. Choudhury and R. Parthasarthy, Physica B **241,** 403 (1998).

[7] S. L. Chaplot, L. Pintschovius, N. Choudhury and R Mittal, Phys. Rev. B **73,** 94308 (2006).

[8] S. L. Chaplot, L. Pintschovius and R. Mittal, Physica B **385,** 150 (2006).

[9] P. Dawson, M. M. Hargreave and G. R. Wilkinson, J. Phys. C **4,** 240 (1971).

[10] F. Gervais, B. Piriou and F. Cabannes, J. Phys. Chem. Solids **34,** 1785 (1973).

[11] K. B. Lyons, M.D. Sturge and M. Greenblatt, Phy. Rev. B **30,** 2127 (1984).

[12] J. H. Nicola and H. N. Rutt, J. Phys. C **7,** 1381 (1973).

[13] R.W. G. Syme, D. J. Lockwood and H. J. Kerr, J. Phys. C **10**, 1335 (1977).

[14] M. P. Lahalle, J. C. Krupa, M. Lepostollec and J. P. Forgerit, J. Solid State Chem. **64**, 181 (1986).

[15] G. M. Rignanese, X. Gonze, G. Jun, K. Cho and A. Pasquarello, Phy. Rev. B **69**, 184301 (2004).

[16] L. Mazeina, S. V. Ushakov, A. Navrotsky and L. A. Boatner, Geochimica et Cosmochimica Acta **69**, 4675 (2005). F. Dachille and R. Roy, J. Geol. **72**, 243 (1964). A. M. Seydoux and J. –M. Montel, EUG IX, Terra Nova 9, Abstract Supplement **1**, 42119 (1997).

[17] B. Manoun, R. T. Downs and S. K. Saxena, American Mineralogist **91,** 1888 (2006).

[18] C. B. Finch, L. A. Harris and C. W. Clark, American Mineralogist **49**, 782 (1964).

[19](a)A. Sen, S. L. Chaplot and R. Mittal, Physica B **363**, 213 (2005); N. Choudhury, S. L. Chaplot, and K. R. Rao, Phy. Rev. B **33**, 8607 (1986); N. Choudhury and S. L. Chaplot, Solid State Comm. **114**, 127 (2000). (b) S. L. Chaplot, Phy. Rev. B **36**, 8471 (1987).

[20]R. Mittal, S. L. Chaplot and N. Choudhury, Progress of Materials Science **51**, 211 (2006).

[21]S. L. Chaplot, N. Choudhury, S. Ghose, Mala N. Rao, R. Mittal and K. N. Prabhatasree, European Journal of Mineralogy **14**, 291 (2002).

[22](a) G. Venkatraman, L. Feldkamp and V. C. Sahni, *Dynamics of Perfect Crystals* (The MIT Press, Cambridge, 1975). (b) P. Bruesch in *Phonons: Theory and Expt*. **Vol. 1** and **Vol. 2** (Springer, Berlin 1982).

[23]S. L. Chaplot (unpublished).

[24](a) www.ncnr.nist.gov. (b) V. F. Sears, Neutron News **3**, 29 (1992). (c) A. -J. Dianoux and G. Lander (Eds.), *Neutron Data Booklet, Institut Laue-Langevin*, Grenoble, France (2002).

[25]M. Taylor and R. C. Ewing, Acta Cryst. B **34,** 1074 (1978).

[26]J. A. Speer and B. J. Cooper, American Mineral. **67**, 804 (1982).

[27]L. H. Fuchs and Elizabeth Gebert, American Mineralogist **43**, 243 (1958).

[28]H. Ozkan, Journal of Applied Physics **47**, 4772 (1976).

[29]H. Ozkan and J. C. Jamieson, Phys. Chem. Miner. **2**, 215 (1978).

[30]S. Ono, Y. Tange, I. Katayama, and T. Kikegawa, American Mineralogist **89**, 185 (2004).

TABLE. I. Comparison between the experimental[25-27] (at 293 K) and calculated structural parameters (at 0 K) of zircon and scheelite phase of $HfSiO_4$ and $USiO_4$ and of zircon and huttonite phase of $ThSiO_4$. For zircon structure (body centered tetragonal, $I4_1/amd$) the M(Hf, Th, U) , Si and O atoms are located at (0, 0.75, 0.125), (0, 0.25, 0.375) and (0, $u$, $v$) respectively and their symmetry equivalent positions are 4a, 4b and 16h respectively. For huttonite structure (monoclinic, $P2_1/n$) the M(Th), Si and O atoms are located at ($u$, $v$, $w$) and their symmetry equivalent positions. For scheelite structure ($I4_1/a$) the M(Hf, U), Si and O atoms are located at (0,0,0.5), (0,0,0) and ($u$,$v$,$w$) respectively.

| | Experimental[26] $HfSiO_4$ (Zircon) | Calculated $HfSiO_4$ (Zircon) | Calculated $HfSiO_4$ (Scheelite) | Experimental[27] $USiO_4$ (Zircon) | Calculated $USiO_4$ (Zircon) | Calculated $USiO_4$ (Scheelite) |
|---|---|---|---|---|---|---|
| $a$ (Å) | 6.57 | 6.48 | 4.68 | 6.981 | 6.76 | 4.88 |
| $c$ (Å) | 5.96 | 6.06 | 10.68 | 6.250 | 6.21 | 11.35 |
| $u$ | 0.0655 | 0.070 | 0.253 | 0.070 | 0.076 | 0.248 |
| $v$ | 0.1948 | 0.207 | 0.146 | 0.222 | 0.212 | 0.128 |
| $w$ | | | 0.070 | | | 0.069 |
| Volume/atom(Å$^3$) | 21.44 | 21.19 | 19.55 | 25.38 | 23.66 | 22.52 |
| Volume/primitive cell(Å$^3$) | 128.63 | 127.14 | 117.28 | 152.29 | 141.93 | 135.10 |

| | Experimental[25] $ThSiO_4$ (Zircon) | Calculated $ThSiO_4$ (Zircon) | Experimental[25] $ThSiO_4$ (Huttonite) | Calculated $ThSiO_4$ (Huttonite) |
|---|---|---|---|---|
| $a$ (Å) | 7.1328 | 6.92 | 6.784 | 6.67 |
| $b$ (Å) | | | 6.974 | 6.83 |
| $c$ (Å) | 6.3188 | 6.31 | 6.500 | 6.63 |
| $u$ | 0.0732 | 0.079 | | |
| $v$ | 0.2104 | 0.215 | | |
| $\beta$ | | | 104.92 | 105.8 |
| Volume/atom(Å$^3$) | 26.79 | 25.16 | 24.76 | 24.21 |
| Volume/primitive cell(Å$^3$) | 160.74 | 150.95 | 297.16 | 290.46 |

| | Experimental[25] ($ThSiO_4$) (Huttonite) | | | Calculated ($ThSiO_4$) (Huttonite) | | |
|---|---|---|---|---|---|---|
| | $u$ | $v$ | $w$ | $u$ | $v$ | $w$ |
| M | 0.2828 | 0.1550 | 0.0988 | 0.282 | 0.156 | 0.089 |
| Si | 0.3020 | 0.1616 | 0.6117 | 0.303 | 0.157 | 0.608 |
| O1 | 0.3900 | 0.3388 | 0.4967 | 0.392 | 0.329 | 0.509 |
| O2 | 0.4803 | 0.1060 | 0.8234 | 0.473 | 0.098 | 0.806 |
| O3 | 0.1216 | 0.2122 | 0.7245 | 0.129 | 0.209 | 0.707 |
| O4 | 0.2451 | 0.4976 | 0.0626 | 0.240 | 0.502 | 0.065 |

TABLE. II. The elastic constants and bulk modulus in zircon and scheelite phase of $ZrSiO_4$ and $HfSiO_4$ and zircon phase of $ThSiO_4$ and $USiO_4$ (in GPa units).

| Elastic constant | Expt.[#] $ZrSiO_4$ (Zircon) | Calc.[7] $ZrSiO_4$ (Zircon) | Calc.[7] $ZrSiO_4$ (Scheelite) | Calc. $HfSiO_4$ (Zircon) | Calc. $HfSiO_4$ (Scheelite) | Calc. $ThSiO_4$ (Zircon) | Calc. $USiO_4$ (Zircon) |
|---|---|---|---|---|---|---|---|
| $C_{11}$ | 424.4 | 432 | 470 | 441 | 477 | 334 | 370 |
| $C_{33}$ | 489.6 | 532 | 288 | 537 | 282 | 453 | 483 |
| $C_{44}$ | 113.3 | 110 | 74 | 107 | 72 | 78 | 89 |
| $C_{66}$ | 48.2 | 39 | 133 | 41 | 136 | 11 | 20 |
| $C_{12}$ | 69.2 | 73 | 241 | 77 | 247 | 38 | 48 |
| $C_{13}$ | 150.2 | 180 | 255 | 192 | 274 | 144 | 159 |
| B | 205[#] | 251 | 303 | 260 | 314 | 197 | 217 |

[#] Expt. Ref. 29 for elastic constants and Ref. 30 for Bulk modulus for zircon.

FIG. 1. (Color online) Ball and stick representation of the zircon (space group $I4_1/amd$) and scheelite ($I4_1/a$) and huttonite ($P2_1/n$) phases of MSiO$_4$ (M=Hf, Th, U). SiO$_4$ tetrahedra are also shown. The solid circles denote M, Si, and O atoms in decreasing order of size.

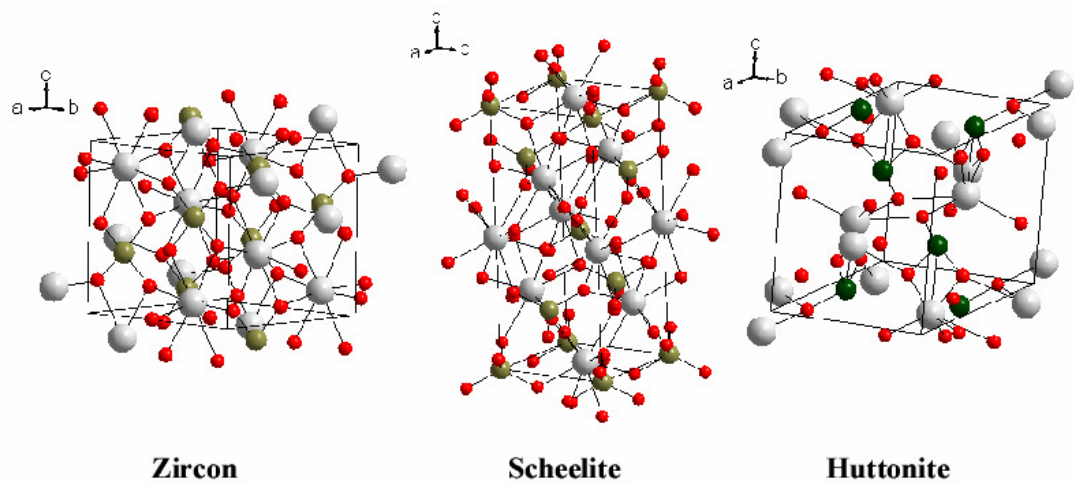

FIG. 2. (Color online) The comparison between the calculated (T=0 K) and experimental[9-14] (T=300 K) zone center phonon frequencies for zircon phase of $MSiO_4$ (M=Zr, Hf, Th, U). The *ab-initio* calculations[15] for $ZrSiO_4$ and $HfSiO_4$ are also shown. The $A_{2g}$, $A_{1u}$, $B_{1u}$ and $B_{2u}$ are optically inactive modes. The frequencies are plotted in the order of $ZrSiO_4$, $HfSiO_4$, $ThSiO_4$ and $USiO_4$ from below.

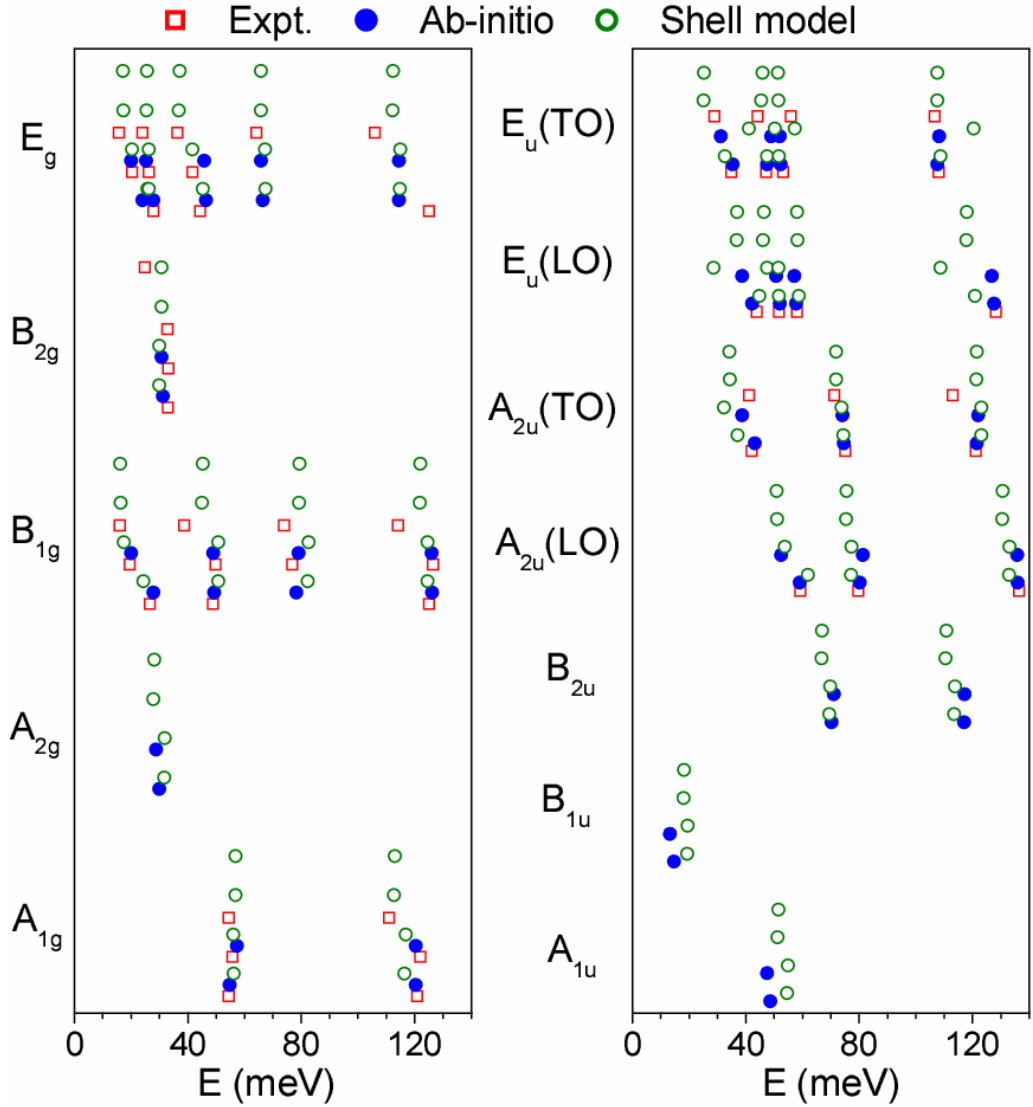

FIG. 3. Comparison of phonon dispersion relations along the high symmetry direction Λ [001] in the ambient as well as high pressure phases of MSiO$_4$ (M=Hf, Th) as calculated at P=0.

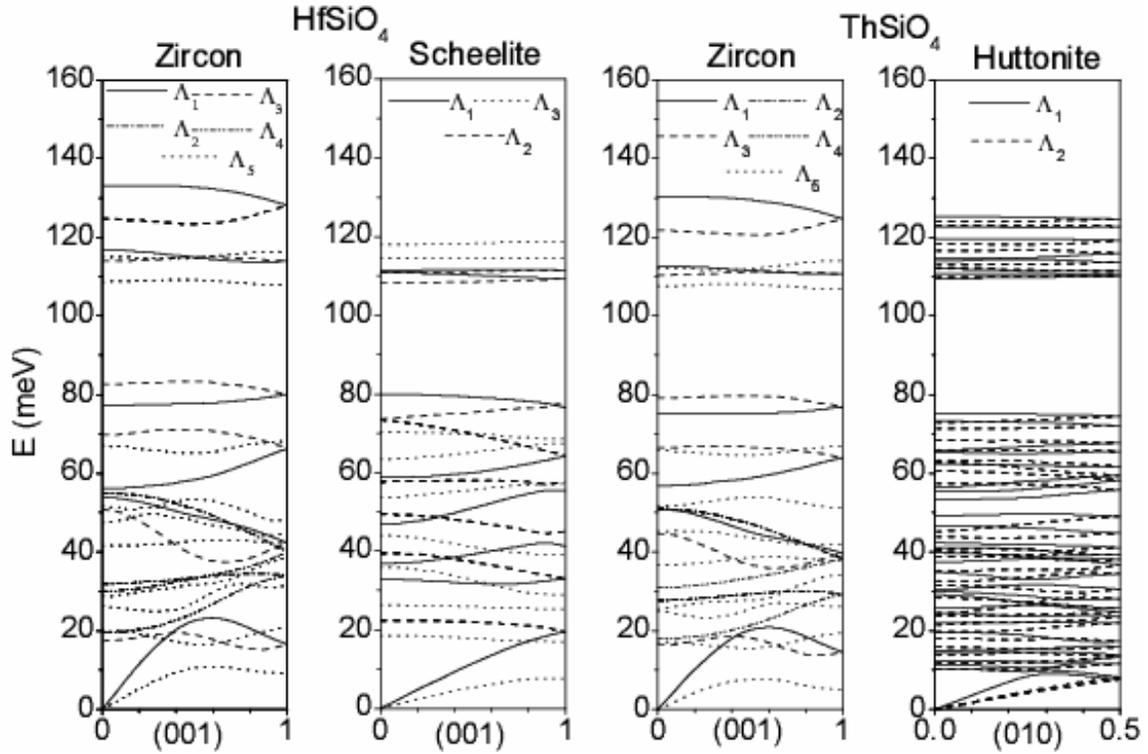

FIG. 4. (Color online) The calculated partial density of states of various atoms and the total density of states of MSiO$_4$ (M=Hf, Th, U) at P=0.

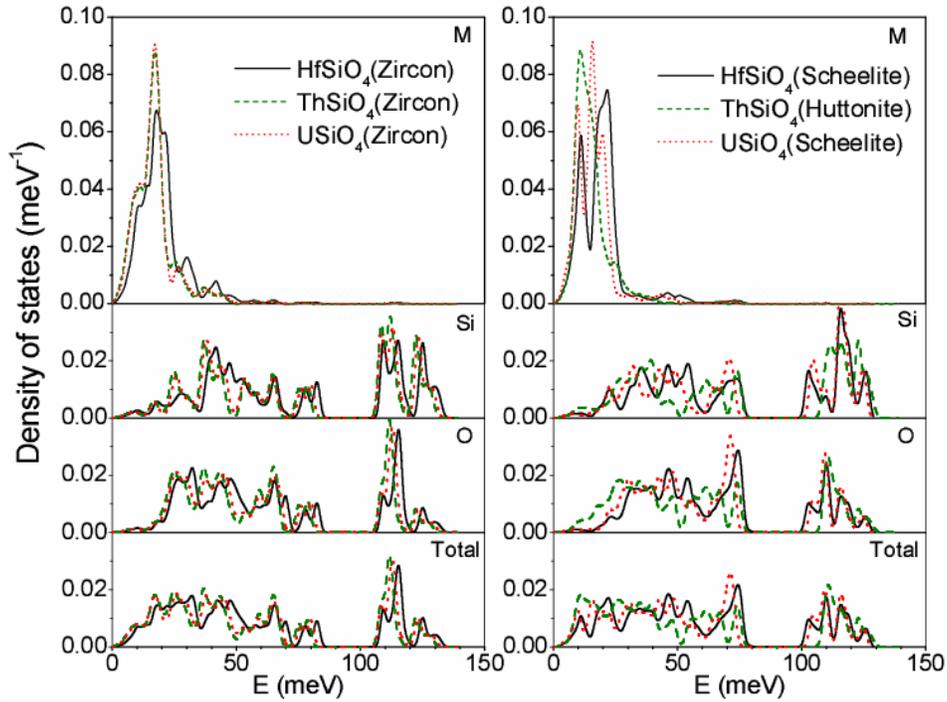

FIG. 5. The calculated neutron-cross-section-weighted phonon density of states in MSiO$_4$ (M=Hf, Th, U) in various phases at P=0.

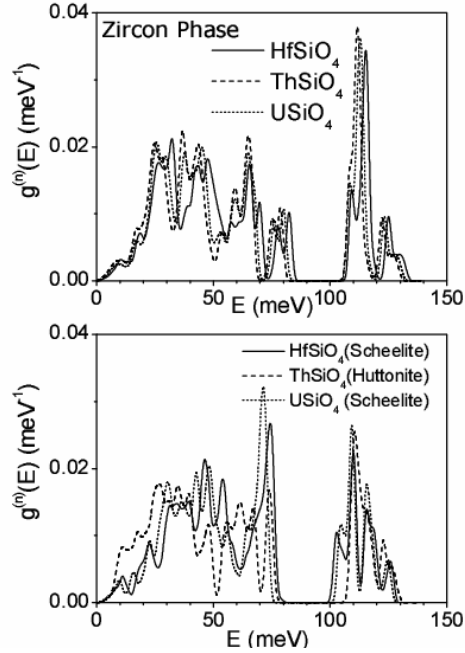

FIG. 6. The calculated specific heat in the ambient as well as high pressure phases of MSiO$_4$ (M=Hf, Th, U) at P=0.

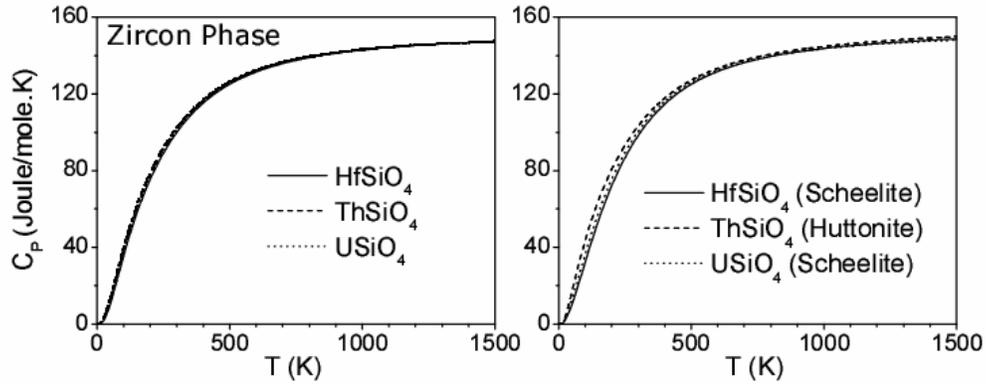

FIG. 7. The calculated Grüneisen parameter as a function of MSiO$_4$ (M= Hf, Th, U) at P=0.

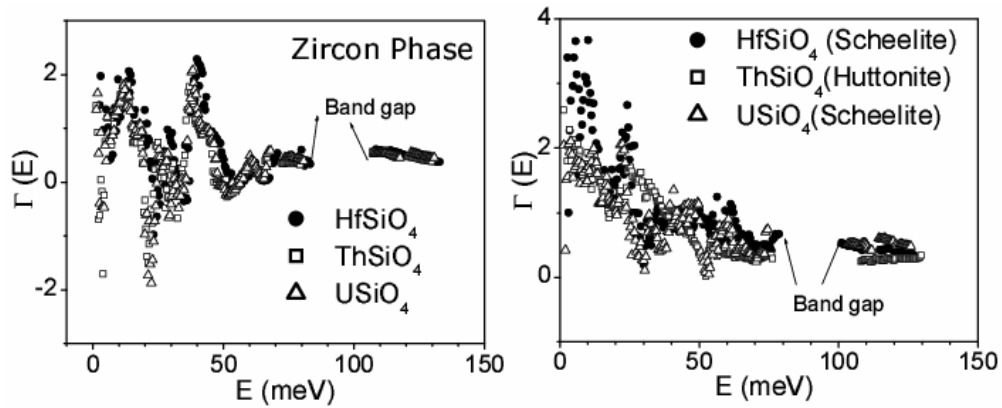

FIG. 8. The calculated thermal expansion behavior of $MSiO_4$ (M= Hf, Th, U) at P=0.

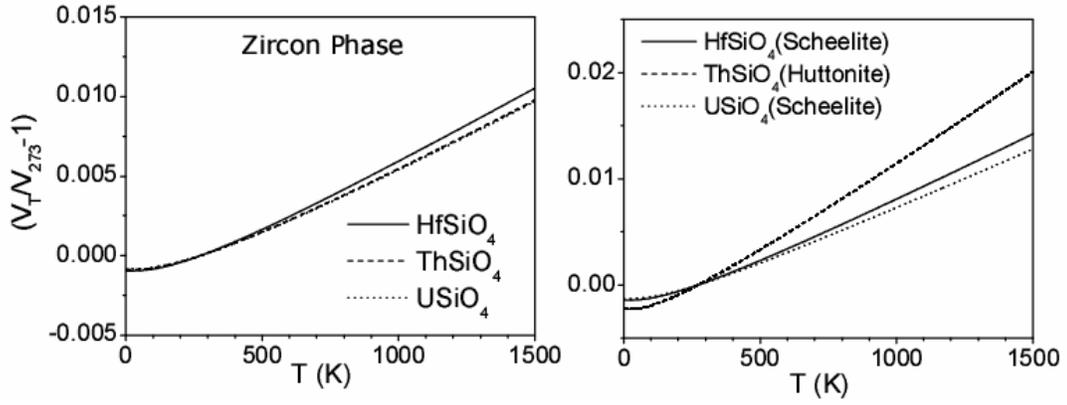

FIG. 9. The calculated equation of state of $MSiO_4$ (M= Hf, Th, U) at T=0. $X_p(a_p,c_p,V_p)$ and $X_o(a_o,c_o,V_o)$ refer to the values at pressure P and ambient pressure, respectively.

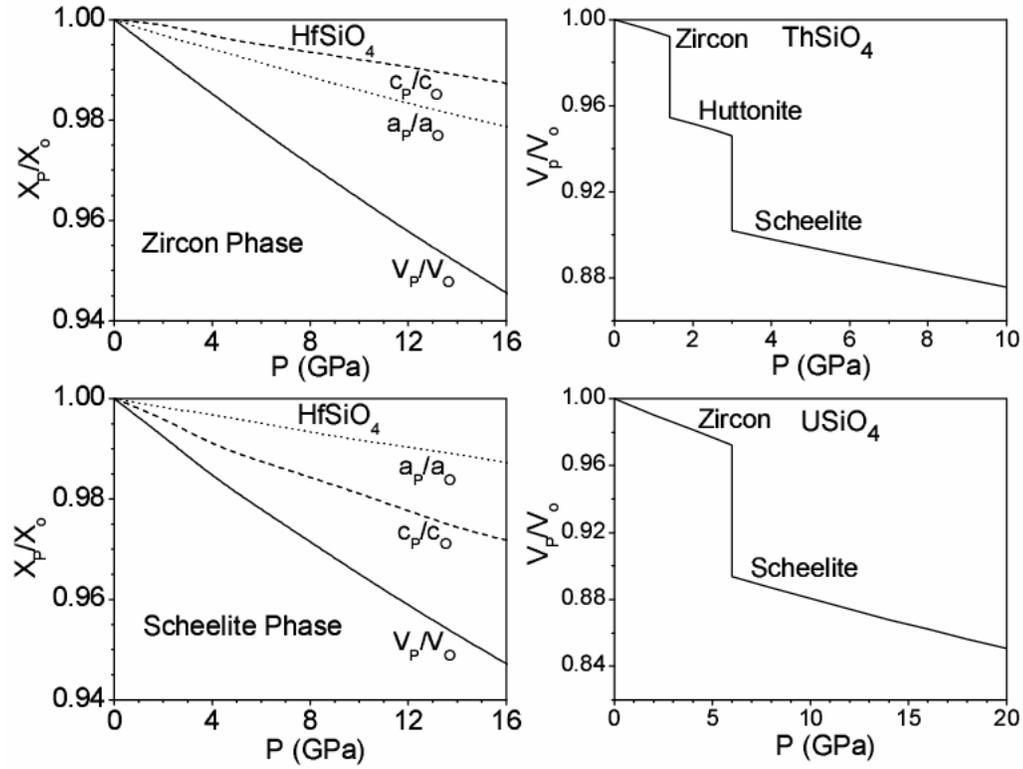

FIG. 10. Typical plots of the differences in the free energies of competing phases in $MSiO_4$ as a function of pressure or temperature.

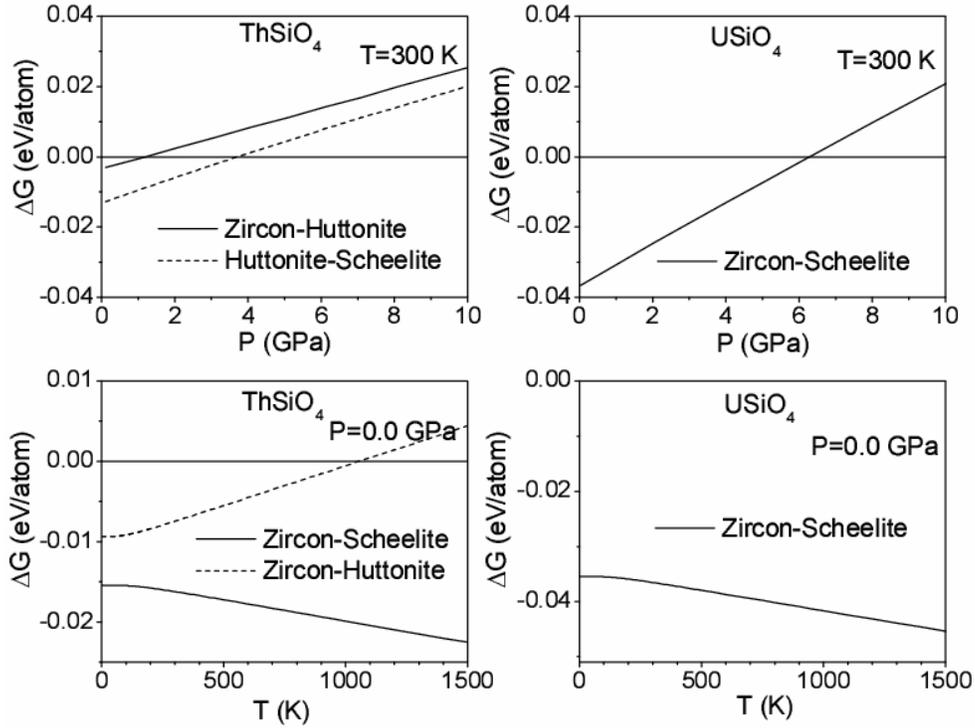

FIG. 11. (Color online) The calculated phase diagram of $HfSiO_4$, $ThSiO_4$ and $USiO_4$ as obtained from the free energy calculations. For $ThSiO_4$, symbols are the experimental data taken from Dachille and Roy.[16] The experimental data for $HfSiO_4$ are from Ref. [17].

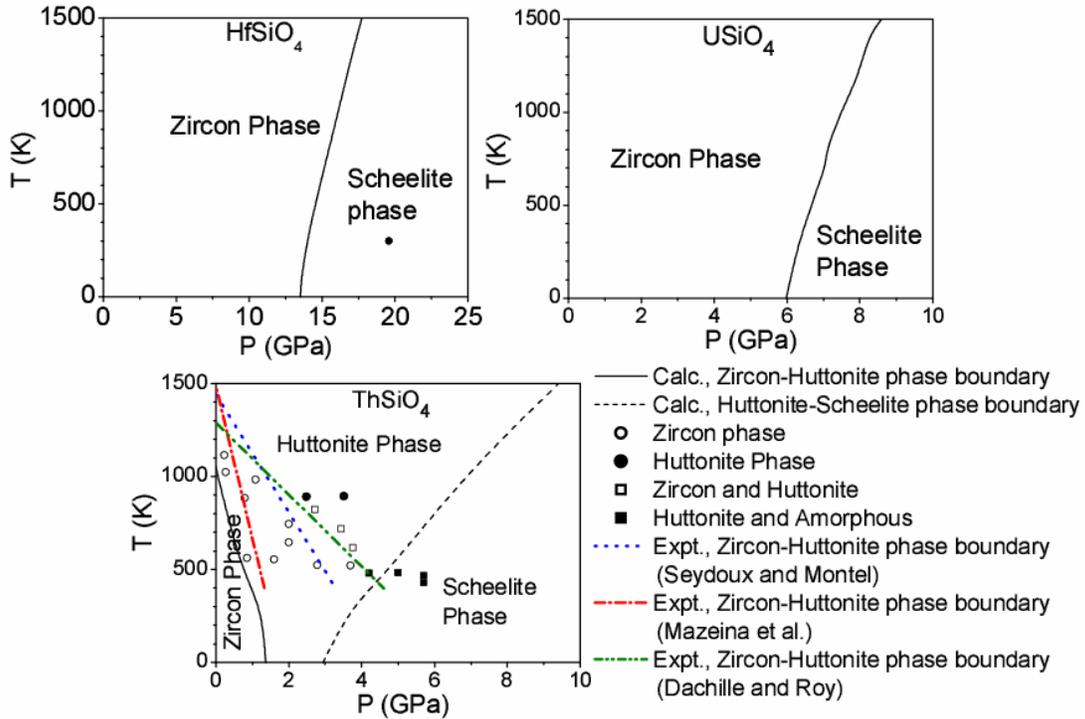

FIG. 12. (Color online) The calculated contribution to the mean squared amplitude of various atoms arising from phonons of energy E (integrated over the Brillouin zone) at T=300 K in various phases of ThSiO$_4$. The atoms are labeled as indicated in Table I.

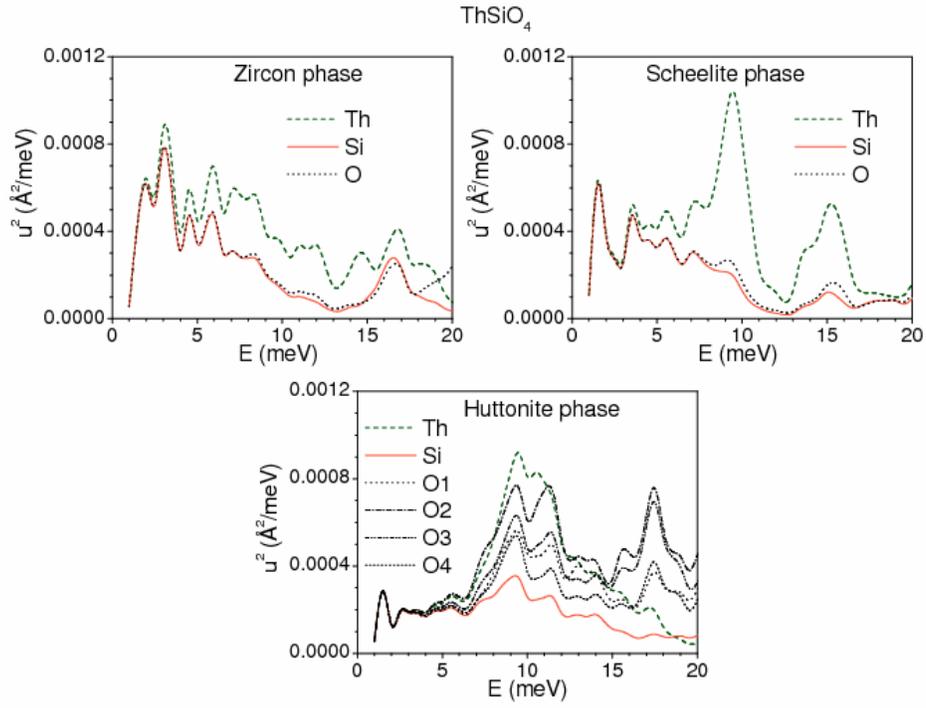